\documentclass[sigconf]{acmart}
\pdfoutput=1
\usepackage{amsmath,amssymb,amsfonts}
\usepackage{algorithmic}
\usepackage{graphicx}
\usepackage{textcomp}
\usepackage{xcolor}

\usepackage{amsthm}
\usepackage{booktabs} 
\usepackage{adjustbox}
\usepackage{enumitem}
\usepackage{subcaption} 
\usepackage{multirow} 
\theoremstyle{definition}
\newtheorem{definition}{Definition}
\DeclareMathOperator*{\argmax}{arg\,max}

\def\model{\textit{Job2Skills}}
\newcommand{\cut}[1]{}

\usepackage{comment}

\newcommand\hl[1]{#1} 

\setcopyright{acmcopyright}

\acmDOI{10.475/123_4}

\acmISBN{123-4567-24-567/08/06}

\acmConference[KDD'20]{ACM SIGKDD Conference on Knowledge Discovery and Data Mining}{August 2020}{San Diego, California USA}
\acmYear{2020}
\copyrightyear{2020}

\acmPrice{15.00}

\begin{document}
\title{Salience and Market-aware Skill Extraction for Job Targeting}

\author{Baoxu Shi}
\affiliation{%
  \institution{LinkedIn Corporation, USA}
}
\email{dashi@linkedin.com}

\author{Jaewon Yang}
\affiliation{%
  \institution{LinkedIn Corporation, USA}
}
\email{jeyang@linkedin.com}

\author{Feng Guo}
\affiliation{%
  \institution{LinkedIn Corporation, USA}
}
\email{feguo@linkedin.com}

\author{Qi He}
\affiliation{%
  \institution{LinkedIn Corporation, USA}
}
\email{qhe@linkedin.com}

\renewcommand{\shortauthors}{B. Shi et al.}

\begin{abstract}
At LinkedIn, we want to create economic opportunity for everyone in the global workforce. 
To make this happen, LinkedIn offers a reactive Job Search system, and a proactive Jobs You May Be Interested In (JYMBII) system to match the best candidates with their dream jobs. 
One of the most challenging tasks for developing these systems is to properly extract important skill entities from job postings and then target members with matched attributes. 
In this work, we show that the commonly used text-based \emph{salience and market-agnostic} skill extraction approach is sub-optimal because it only considers skill mention and ignores the salient level of a skill and its market dynamics, \textit{i.e.}, the market supply and demand influence on the importance of skills. 
To address the above drawbacks, we present \model, our deployed \emph{salience and market-aware} skill extraction system. 
The proposed \model ~shows promising results in improving the online performance of job recommendation (JYMBII) ($+1.92\%$ job apply) and skill suggestions for job posters ($-37\%$ suggestion rejection rate). 
Lastly, we present case studies to show interesting insights that contrast traditional skill recognition method and the proposed \model~from occupation, industry, country, and individual skill levels.
Based on the above promising results, we deployed the \model ~online to extract job targeting skills for all $20$M job postings served at LinkedIn.

\end{abstract}


\maketitle

\section{Introduction}\label{sec:introduction}

LinkedIn is the world's largest professional network whose vision is to ``\textit{create economic opportunity for every member of the global workforce}''. To achieve this vision, it is crucial for LinkedIn to match {job postings to \emph{quality applicants} who are both qualified and willing to apply for the job.}
To serve this goal, LinkedIn offers two job targeting systems, namely Jobs You May Be Interested In (JYMBII)~\cite{gu2016learning}, which proactively target jobs to the quality applicants, and Job Search~\cite{li2016get}, which reactively recommend jobs that the job seeker qualifies.

Matching jobs to the quality applicants is a challenging task. Due to high cardinality (for LinkedIn, $645$M members and more than $20$M open jobs), it is computationally intractable to define job's target member set by specifying individual members. For this reason, many models match candidates by profile attributes~\cite{volkovs2017content,paparrizos2011machine,abel2015we}. At LinkedIn, we mostly use titles and skills to target candidates (members). In other words, when a member comes to the job recommendation page, we recommend job postings whose targeting skills (or titles) match the member's skills (or titles).

\begin{figure}[t]
    \centering
    \begin{adjustbox}{max width=.95\linewidth}
        \includegraphics[width=\textwidth]{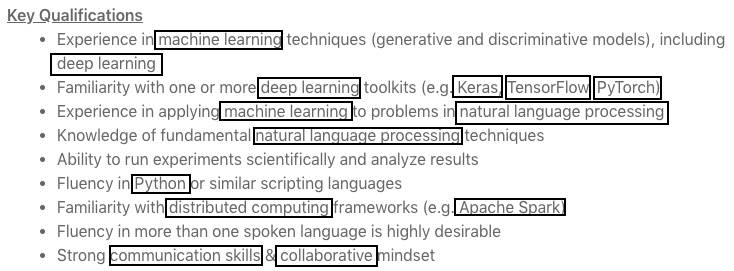}
    \end{adjustbox}
    \caption{Snippet of a machine learning engineer job posted on LinkedIn. Squared text are detected skill mentions.}
    \label{fig:key_qualifications}
\end{figure}

In this paper, we study how to target job postings by identifying relevant skills. Given a job posting, we extract skills from the job posting so that it can be shown to the members who have such skills. This task of mapping the job to a set of skills is very important because it determines the quality of applications for the job posting and affects hiring efficiency.

To improve the quality of applicants, what should be the objectives of extracting skills for job targeting? We argue that the extracted skills need to meet two criteria. First, the extracted skills should not only be mentioned in the job posting, but also be relevant to the core job function. In other words, the skills should be \emph{salient} to the job posting. Second, the extracted skills should be able to reach out to enough number of members. In other words, there should be enough supply for the skills in the job market. In summary, we aim to build a machine learning model to extract skills from job postings in a \emph{salience and market-aware} way.

However, developing such salience and market-aware skill extraction model is a very challenging task. {Not only because modeling the salience and market dynamic is hard, but also because the lack of ground-truth data}. Given a job posting, it is tricky to get gold-standard labels for the skills that are both salient and have strong supply. One may propose using crowdsourcing to annotate job postings, but there are two problems. First, crowdsourcing may not be the {most cost-efficient way} to collect a large amount of training examples. Second and more importantly, labeling salient and high-supply skills requires very solid domain knowledge about the job market and the job posting itself. As we will show later, the labels collected by crowdsourcing do not give us the most salient skills with the best job market supply. 

Figure~\ref{fig:key_qualifications} gives an illustration of the above challenge. Given a job posting, it is relatively straightforward for human to label skills mentioned in the posting (rectangles), and an annotator in the crowdsourcing platform would be able to do it. However, if the annotator aims to label which skills are more salient and have better job market supply, the annotator needs to understand both the job market and the job description well. For example, for the sake of market supply, the annotator should choose ``deep learning'' instead of specific tools such as ``Pytorch'' or ``Keras'', but this is impossible if the annotator does not understand deep learning related skills and the deep learning job market. On the other hand, skills like ``Communication Skill'' have large supply but they are less salient in the context of deep learning engineer jobs.

Previous works usually treat this skill extraction task as a named entity recognition (NER)~\cite{nadeau2007survey} task and use the standard named entity recognizer to identify skill mentions~\cite{vasudevan2018estimating, li2016get}. In the example of Fig.~\ref{fig:key_qualifications}, existing methods would identify all rectangles and use them for job targeting with equal importance. This would lead to showing this deep learning job to members who have ``Communication Skills''. Since these methods do not consider the job market supply and salience of entities, they will lead to sub-optimal job targeting performance, as we will show later.

\noindent{}\textbf{Present work: Data collection.} In this paper, we tackle the problem of building salience and market-aware skill extraction model for job targeting. To collect the ground-truth labels, we note that the ground-truth skills need to meet dual criteria: salience and market-awareness. Since it is hard to come up with one data collection method to match both criteria, our strategy is to develop one method to cover each criterion, and combine the data collected using two methods. First option is focused on salience. We ask hiring experts to pick which skills they would want to use for targeting of their job postings. Since this method is from the salience perspective, we also conducted analysis to validate that the collected labels match our notions of market-aware skill extraction. This option allows us to collect explicit feedback from hiring experts, but this option is only available to a small portion of job postings that are created through the LinkedIn's job creation flow.

To cover market-awareness and increase the training data size, we apply ``distant supervision'' to get weak labels from other job postings. In particular, for a given job posting, we track which members applied for the job and received positive feedback. We then identify common skills that these successful applicants have and use these skills as the ground-truth skills. We then perform analysis to validate that these skills are also salient.

\noindent{}\textbf{Present work: Modeling job market and entity salience.} After collecting the ground-truth, we develop \model, a novel machine learning model to extract skills in a market-aware, salience-aware way. We note that retraining existing NER models leads to sub-optimal results as they rely purely on the text information, ignoring important signals such as how much supply the skill has in the job market, how salient the skill is overall, and so on. Therefore, we add signals representing the job market supplies and the salience of skills in the model and significantly improve the performance in the offline evaluation and online A/B test.

\noindent{}\textbf{Present work: Product improvements.} We deploy the \model~ in production and improve the product metrics across multiple applications. First, we ask feedback from hiring experts in the job creation systems and outperform the existing NER-based model by more than $30\%$. {Second, we employ this new skill extraction model in the job recommendation systems, one of the most important recommender systems at LinkedIn. We observe $2\%$ member-job interaction and $6\%$ coverage improvement in proactive job recommender (JYMBII) system.} 

\noindent{}\textbf{Present work: Insights for Job Market.} 
We note that \model's outputs capture the hiring trend in the job market for millions of companies that post jobs at LinkedIn. Since \model~is trained on the feedback from hiring experts and job markets, it learns what kinds of talent that each employer is trying to recruit. We argue that \model~reveals employers' intention better than traditional information extraction methods that do not consider salience and market factors. 
We present extensive studies to showcase the power of insights that can be gained with \model. For example, we show that \model's results can forecast that Macy's would expand a tech office in a new location in two months before the official news article comes out. We also show that \model~ can vividly capture the rising and fall of popularity of a certain skill. Lastly we demonstrate that \model~ can be used to compare the required skill sets across different regional markets.

The contributions of this paper are summarized as follow:
\begin{itemize}
\item We propose a skill extraction framework to target job postings by skill salience and market-awareness, which is different from traditional entity recognition based method.
\item We devise a data collection strategy that combines supervision from experts and distant supervision based on {massive job market interaction history}.
\item We develop \model ~by combining NER methods with {additional market and deep learning powered salience signals}.
\item We deploy \model ~to LinkedIn to improve overall hiring efficiency across multiple products.
\item We provide a case study to show how the proposed market-aware skill extraction model yields better skill-related insights about the workforce and beyond.
\end{itemize}

\section{Problem Definition}~\label{sec:problem_definition}

Here we will provide the formal definition of the job targeting task and show how we utilize the job targeting process to formulate our salience and market-aware skill extraction task.

\begin{definition}\label{def:job_targeting}
\textbf{Job targeting} is an optimization task where given job posting $p$ and member set $\mathbf{M}$, find a $t_m$-sized target candidate set $\mathbf{M}_p\subset\mathbf{M}$ of $p$ that can maximize the probability that member $m\in\mathbf{M}_p$ belongs to the job posting $p$'s quality applicants set $\mathbf{A}_p$, {a set of members who are likely to apply for the job and get positive feedback from the job posters}.
\end{definition}

We can write the objective of Def.~\ref{def:job_targeting} as

\small
\begin{equation}\label{eq:job_targeting}
\argmax_{\mathbf{M}_p \subset \mathbf{M}, |\mathbf{M}_p|=t_m}\sum_{m\in \mathbf{M}_p}\mathsf{Pr}(m\in\mathbf{A}_p).
\end{equation}
\normalsize

This optimization is intractable because it involves combinatorial optimization for $645$ million LinkedIn members. For this reason, existing models tend to build job targeting models with attributes instead of directly targeting individuals. \hl{Moreover, modeling via attributes also makes the model human-interpretable.} LinkedIn, for example, uses attribute entities such as job title, company, screening question~\cite{shi2020job2questions}, and skill to target members \hl{and provide interpretable insights}~\cite{shan2020jobstandardization}. Among these entities, we discovered that skill is the most critical entity type for the job targeting task.

Therefore, we formulate a problem of skill-based job targeting. Given a job posting $p$, we choose $t_s$-sized job targeting skills $\mathbf{S}_p$ and we show the job postings to member $m$ if $m$ has \textit{at least} one matching skill in their skill set $\mathbf{S}_m$. Since members declare their skill sets $\mathbf{S}_m$ on their profile, we need to identify $\mathbf{S}_p$ only. Our objective becomes:

\small
\begin{equation}\label{eq:job_targeting_2}
\argmax_{\mathbf{S}_p \subset \mathbf{S}, |\mathbf{S}_p|=t_s}\sum_{m\in\mathbf{M}}\mathbb{I}(\mathbf{S}_m\cap\mathbf{S}_p\neq\emptyset)\mathsf{Pr}(m\in\mathbf{A}_p),
\end{equation}
\normalsize

\noindent{}{where $\mathbb{I}(\cdot)$ is an indicator function and $\mathbf{S}$ is the full skill entity set used by LinkedIn. 
Although Eq.~\ref{eq:job_targeting_2} reduces the dimensionaliy of search space from $645$ million members to $40$ thousand skills, it is still combinatorial optimization and thus very hard to optimize directly. 
To make it more tractable, we make the following assumptions. We assume that each skill $s$ has a utility $\mathsf{U}(m,p,s)$ that is the chance of $m$ being qualified for $p$ when $m$ knows skill $s$, and that the probability of being qualified $\mathsf{Pr}(m\in\mathbf{A}_p)$ is the sum of utility $\mathsf{U}(m,p,s)$ of each skill $s \in \mathbf{S}_m\cap\mathbf{S}_p $. In other words, we assume that each skill $s$ increases the chance of $m$ being qualified by $\mathsf{U}(m,p,s)$.
}
\small
\begin{equation}\label{eq:job_targeting_approx}
\mathsf{Pr}(m\in\mathbf{A}_p) \approx \sum_{s\in\mathbf{S}_p}\mathbb{I}(s\in\mathbf{S}_m)\mathsf{U}(m,p,s).
\end{equation}
\normalsize

\noindent{}{
With this notation, Eq.~\ref{eq:job_targeting_2} becomes}

\small
\begin{equation}\label{eq:job_targeting_3}
\argmax_{\mathbf{S}_p \subset \mathbf{S}, |\mathbf{S}_p|=t_s}\sum_{s\in\mathbf{S}_p}\sum_{m\in\mathbf{M}}\mathbb{I}(s\in\mathbf{S}_m)\mathsf{U}(m,p,s) = \sum_{s\in\mathbf{S}_p}\sum_{m\in\mathbf{M}_s}\mathsf{U}(m,p,s).
\end{equation}
\normalsize

\noindent{}{where $\mathbf{M}_s$ is members who have $s$ in their skill set.
We can further simplify the notation by introducing $\mathsf{U}(p, s) = \sum_{m\in\mathbf{M}_s}\mathsf{U}(m,p,s)$:
}

\small
\begin{equation}\label{eq:job_targeting_final}
\argmax_{\mathbf{S}_p \subset \mathbf{S}, |\mathbf{S}_p|=t_s}\sum_{s\in\mathbf{S}_p}\mathsf{U}(p, s),
\end{equation}
\normalsize

\noindent{}{where $\mathsf{U}(p, s)$ is the sum of $\mathsf{U}(m,p,s)$ for all members who have $s$. Given that $\mathsf{U}(m,p,s)$ is the increase in probability of $m$ being qualified for $p$ by knowing $s$, $\mathsf{U}(p, s)$ quantifies the overall increase in qualified applicants by targeting members having $s$. We call $\mathsf{U}(p, s)$ the skill $s$'s utility for job posting $p$.
}

{
The formulation in Eq.~\ref{eq:job_targeting_final} makes optimization much more tractable and simpler than the original form in Eq.~\ref{eq:job_targeting}. In Eq.~\ref{eq:job_targeting_final}, we optimize the sum of utilities $\mathsf{U}(p, s)$ for each skill $s$. Therefore, choosing the optimal $\mathbf{S}_p$ can be done by picking $t_s$ skills that have the highest value of $\mathsf{U}(p, s)$ for a given job posting $p$.
We call this problem \emph{salience and market-aware skill extraction}. We name the problem in this way because we find that the skill $s$ needs to satisfy the following two criteria to have high utility $\mathsf{U}(p, s)$.
}

\noindent{1. \textbf{Skills should have sufficient market supply}.} {
In order for $\mathsf{U}(p, s)$ to be high, the size of $\mathbf{M}_s$ needs to be large enough. In other words, there must be a sufficient number of members that have skill $s$.
}

\noindent{2. \textbf{Skills should be salient to the job posting}.} {
Another factor that determines $\mathsf{U}(p, s)$ is the value of $\mathsf{U}(m,p,s)$. Remember that $\mathsf{U}(m,p,s)$ quantifies the chance that the member $m$ is qualified for job posting $p$ if the member has skill $s$. If $s$ is a core, ``salient'' skill for the job posting $p$, it will have high $\mathsf{U}(m,p,s)$.
}

{
Combining the above two criteria together, we define the salience and market-aware skill extraction task as:
}

\begin{definition}\label{def:market_aware}
\textbf{Salience and Market-aware skill extraction} is an optimization task where given a job posting $p$ and skill set $\mathbf{S}$, we estimate the utility $\mathsf{U}(s,p)$ that is the increase in qualified applicants by targeting members with $s$.
\end{definition}

{
As a comparison, we discuss the \emph{salience- and market-agnostic approach} that chooses $\mathbf{S}_p$ by identifying the skills mentioned in the job posting content. We can think of it as solving Eq.~\ref{eq:job_targeting_final} with modified utility $\mathsf{U'}(p, s)$ that purely depends on whether the skill is mentioned in the job posting. We can formally define it as follows:
}

\begin{definition}\label{def:market_agnostic}
\noindent{\textbf{Salience- and market-agnostic skill extraction}} is an optimization task where given a job posting $p$ and skill set $\mathbf{S}$, extract $t_s$-sized skill set $\mathbf{S}_p \subset \mathbf{S}$ for job posting $p$ that maximizes the utility $\mathsf{U'}(s,p;\theta) \propto \mathsf{Pr}(s|c_p)$, which is the likelihood that skill $s$ is mentioned in $p$'s content $c_p$. 
\end{definition}

As shown in Def.~\ref{def:market_agnostic}, this method simplistically defines a skill of a job posting by calculating the probability that it is mentioned in the job posting, using some named skill-entity recognizer. {Because Def.~\ref{def:market_agnostic} does not consider the skill salience and the market supply, the extracted skills are not for targeting quality applicants. Therefore Def.~\ref{def:market_agnostic} is not solving the job targeting task defined in Def.~\ref{def:job_targeting}}.

{Next, we will discuss the methods we use to gather the ground truth $\mathbf{S}_p$ for training our salience and market-aware skill extraction system}, followed by how we learn the utility function described in Def.~\ref{def:market_aware} for the proposed \model ~model.

\begin{figure}[t] 
\centering
  \begin{subfigure}[b]{\linewidth}
    \centering
    \includegraphics[width=\textwidth]{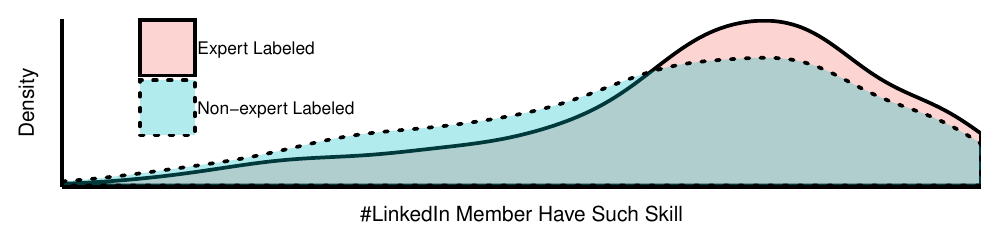}
  \end{subfigure}%
  \caption{Non-expert and expert labeled skills' popularity.} 
\label{fig:skill_distribution} 
\end{figure}

\section{Data Collection}\label{subsec:data_collection}
{As we state before, one of the major challenges of developing a salience and market-aware job targeting system is the lack of ground-truth data. Here we will describes two data collection approaches that address both salience and market-awareness.}

\noindent{\textbf{Collect from Job Posting Experts:}} Although {crowd-sourcing is scalable compared to labeling by group of experts}, {the quality is sub-optimal because non-experts cannot distinguish the level of skill salient and often ignore the market supply.} {Instead of designing a sophisticated skill labeling task for non-experts to collect high-quality data, we directly collect data from experts who select job targeting skills regarding the salience and job market implicitly.}

To do so, we ask job posters to provide job targeting skills. Given a job posting $p$, we collect the job poster saved job targeting skills to form the positive skill set $\mathbf{S}_p^+$, and use the rejected, recommended job targeting skills to create the negative set $\mathbf{S}_p^-$. Lastly, we construct the positive and negative job-skill training pairs $\mathbf{J}^+=\{\langle p, s_{i}\rangle|p_i\in\mathbf{P},s_i\in\mathbf{S}_p^+\}$ and $\mathbf{J}^-=\{\langle p, s_{i}\rangle|p_i\in\mathbf{P},s_i\in\mathbf{S}_p^-\}$, respectively. {We will refer this dataset as the \textit{Job Targeting skill} (JT) dataset.}

{Fig.~\ref{fig:skill_distribution} shows the distribution of crowd-labeled (non-expert) and expert-labeled skills in terms of the number of LinkedIn members having the skill. We observe that expert labeled skills better align with the market supply in terms of member-side skill popularity.
}

\noindent{\textbf{Collect from Market Signals:}}{ Although hiring experts provide high quality data and address both entity salience and market-awareness, such approach is only available to a small portion of job postings that are created through the LinkedIn's job creation flow. To cover more jobs and generate a large amount of labeled data for model training, we decide to get a large amount of weak labels using job market signals instead of human annotations.}

{To be specific, given job posting $p$, we first collect $\mathbf{A'}_{p}$ by assuming quality applicants are the members who apply for job $p$ and receive positive interactions from the recruiters. Note that $\mathbf{A'}_{p}$ is an approximation of the true quality applicant set because we only consider one stage of the recruiting process on LinkedIn. With $\mathbf{A'}_{p}$, we define $\mathbf{S}_p^+=\mathbf{S}_{A'
_{p}}\cap \mathbf{S}_{c_p}$, where $\mathbf{S}_{A'_{p}}$ is a set of common skills shared by $\mathbf{A'}_{p}$, and $\mathbf{S}_{c_p}$ are the skills mentioned in $p$. Similarly, $\mathbf{S}_p^-=\mathbf{S}_{c_p}\setminus \mathbf{S}_{A'_{p}}$, where the negative skills are the ones mentioned in the job posting but not shared by the quality applicants $\mathbf{A'}_{p}$. We will refer this dataset as the \textit{Quality Applicant skill} (QA) dataset.}

{To examine the salient level of these market signal derived skills, we sample jobs with labeled skills from both quality applicants and job targeting datasets, and then compare the top skills shared by quality applicants to the skills labeled by the job posters. We find that the top-ranked quality applicant skills largely overlap with the job poster selections. To be specific, for the top-$5$ quality applicant shared skills, more than $60\%$ of them are labeled as positive skills and $~8\%$ of them are labeled as negative skills by job posters.}

\section{The proposed \model ~model}\label{sec:job2skills}

After we describe the procedure we use to collect the ground truth for salience and market-aware job targeting skill extraction, here we discuss how we {build the proposed} \model {~using multi-resolution skill salience features and market-aware signals.} Compared to simple skill tagging, which merely identifies mentioned skills, multi-resolution skill salience will identify important skills from all mentioned skills.

\subsection{Multi-resolution Skill Salience Feature}

In this work, we hypothesize that good job targeting skills should be salient to the job posting. Unlike other text, job postings are usually long text with several well-structured segments, \textit{e.g.}, requirements, company summary, benefits, etc. To accurately estimate the level of skill salience and fully utilize the rich job posting information, one should not only consider the mentioning text of the skill, but also other segments in the job and the entire job posting. 

\begin{figure}[t]
    \centering
    \begin{adjustbox}{max width=\linewidth}
        \includegraphics{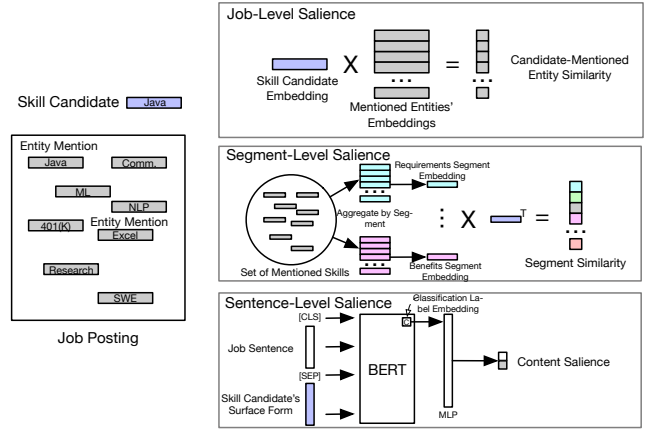}
    \end{adjustbox}
    \caption{{Multi-resolution skill salience estimation.}}
    \label{fig:skill_salience}
\end{figure}

Next, we will first briefly describe how we tag skills from job postings, and then provide details on how we explicitly model the skill salience at three resolution levels: sentence, segment, and job level. Fig.~\ref{fig:skill_salience} gives an overview of the multi-resolution skill salience features we used in this work.

To identify skills from job postings, we first utilize an existing, in-house skill tagger to find out all possible skill mentions. By leveraging a comprehensive skill taxonomy with an extensive set of skill surface forms, the skill tagger can identify the majority of the mentioned skills. We then pass the skill mentions into a feature-based regression model to link them to the corresponding skill entities. After we find all the skills in the job posting, we then model the skill salience from the following three levels:

\noindent{\textbf{Sentence-level  Salience}:}{To estimate the skill salience at sentence level, we build a neural network model to learn the skill salience by modeling the job posting sentence that contains the skill mention and the skill's surface form. The model is defined as}

\begin{equation}
    \mathsf{Pr}(s|c_{\textsf{sent}}) = \sigma(\mathbf{W}\cdot f_{\textsf{encoder}}(c_s, c_{\textsf{sent}}) + b),
\end{equation}

\noindent{}{where $\sigma(x)=1/(1+\exp(-x))$, $s$ is some skill, $c_s$ is the skill's surface form text, $c_{\textsf{sent}}$ is the sentence containing the skill mention, and $f_{\textsf{encoder}}$ is some text to embedding encoder. We tested multiple encoders including FastText~\cite{bojanowski2017enriching}, Universal Sentence Encoder (USE)~\cite{cer2018universal}, and BERT~\cite{devlin2018bert}. For FastText and USE, we encode the skill and sentence separately and then use the concatenated embedding as the encoder output. For BERT, we feed both skill and sentence into the model and pick the embedding of the \textsf{[CLS]} token as the encoder output.}

\noindent{\textbf{Segment-level Salience}:}{ Unlike sentence level skill salience, where the input text length is limited to a single sentence, job segment (one or multiple consecutive paragraphs describing the same topic), \textit{e.g.}, the company summary segment or requirements segment in a job description, are usually longer and much noisier. Therefore, it is not easy to model them directly using the neural network models that are designed for shorter text. Instead of modeling the job segment text directly, we choose to represent a job segment by the embeddings of the skill entities mentioned in the segment. 

To get the entity (title and skill) embeddings~\cite{Recruiter2018LinkedIn, shi2019representation}, we learn LinkedIn's unsupervised entity embeddings using LinkedIn member profile}. The skip-gram loss is defined as:

\small
\begin{equation}\label{eq:loss}
\begin{split}
    \mathcal{L} =& \sum_{m \in \mathbf{M}} \sum_{e\in\mathbf{E}_m}\sum^{k_{\textsf{pos}}}\mathbb{E}_{e_+ \sim \mathsf{Unif}(\mathbf{E}_m \setminus \{e\})}\bigg(\log(\sigma(\mathbf{W}_{e_+}\cdot\mathbf{W}_{e}^T))\\
    &+ \sum^{k_\textsf{neg}}\mathbb{E}_{e_{-}\sim\mathsf{Unif}(\mathbf{E} \setminus \mathbf{E}_m)}\log(\sigma(-\mathbf{W}_{e_{-}}\cdot\mathbf{W}_{e}^T))\bigg),
\end{split}
\end{equation}
\normalsize

\noindent{}{Where $k_{\textsf{pos}}$ and $k_{\textsf{neg}}$ are the number of positive and negative samples, and $\textsf{Unif}$ is a uniform sampling function. Simply put, for each LinkedIn member $m$'s profile entity $e \in \mathbf{E}_m$, we optimize the entity embedding $\mathbf{W}_{e}$ so that the skills and titles appear in the same member's profile entity set $\mathbf{E}_m$ are similar to each other comparing to a random entity $e\in\mathbf{E}$ that is not in $\mathbf{E}_m$. 

After we obtain the entity embedding matrix $\mathbf{W}$, we define the segment-level skill salience as}

\small
\begin{equation}
    \mathsf{Pr}(s|\mathbf{S}_{c_{p,l}}) = \sigma(f_\textsf{meanpool}(\{\mathbf{W}_{s_j} | s_j \in \mathbf{S}_{c_{p,l}}\})\cdot\mathbf{W}_{s}^T),
\end{equation}
\normalsize

\noindent{}{where $\mathbf{S}_{c_{p,l}}$ is set of skills mentioned in the job content $c_p$ with segment label $l$. This measures the similarity between the given skill $s$ and the centroid of other skills mentioned in segment $l$.}

\noindent{\textbf{Job-level Salience}:}{ Similar to segment level salience, Job level skill salience is modeled by the average embedding similarity between the given skill and all entities mentioned in the job posting, which includes skill and title entities.} The salience score is defined as:

\begin{equation}
    \mathsf{Pr}(s|\mathbf{E}_{c_{p}}) = \frac{1}{|\mathbf{E}_{c_{p}}|}\sum_{e\in\mathbf{E}_{c_{p}}}\sigma(\mathbf{W}_s \cdot \mathbf{W}_e^T)
\end{equation}

\noindent{}where $\mathbf{E}_{c_{p}}$ is a set of title and skill entities mentioned in job $p$. Note that we choose to compute entity-wise similarity instead of meanpooled-similarity here because 1) the size of $\mathbf{E}_{c_{p}}$ is often significantly larger than the size of $S_{c_{p,l}}$ and may contain more noisy data, and 2) $\mathbf{E}_{c_{p}}$ contains different type of entities from multiple aspects of the job posting. Based on above observation, we believe it is sub-optimal to compute job-level salience by forming a single meanpooled centroid using all mentioned entities.

In sum, for a skill candidate $s$ of a job posting, we use the above methods to compute multi-resolution salience probability scores ($\mathsf{Pr}(s|c_{\textsf{sent}})$, $\mathsf{Pr}(s|\mathbf{S}_{c_{p,l}})$, and $\mathsf{Pr}(s|\mathbf{E}_{c_{p}})$) as salience features, combine them with the market-aware signals which we will cover in the next section, and then build the final salience and market-aware skill extraction model.

\subsection{Market-aware Signals}

{Besides the salient level of entities, we also hypothesize that good job targeting skills should have sufficient market supply. To model the supply of skills, it is necessary to factor in market-related signals into the proposed} \model{. In general, the market-related signals can be derived from LinkedIn's member base $\mathbf{M}$ and job postings $\mathbf{P}$. Next, we will describe the signals from these two groups.}

\noindent{\textbf{Member Features}:} The goal of the member feature group is to capture how skills can reach a broader audience by measuring member-side skill supply, and therefore to improve job exposure. Here we consider both the general skill supply which measures the overall skill popularity among all members $\mathsf{Pr}(s|\mathbf{M})$, and the cohort affinities, which indicate the skill supply with finer granularity. To be specific, we partition the member set $\mathbf{M}$ into a group of non-empty member subsets (cohorts) $\mathcal{M}$ using different strategies, then compute the point-wise mutual information (PMI) and the entropy (H) of $s$ given a cohort $\mathbf{M}_i$ as follows:

\small
\begin{equation}
    \mathsf{PMI}(\mathbf{M}_i;s) = \log\frac{\mathsf{Pr}(s|\mathbf{M}_i)}{\mathsf{Pr}(s|\mathbf{M})}, H(s)=-\sum_{\mathbf{M}_i\in\mathcal{M}}\frac{\mathsf{Pr}(s|\mathbf{M}_i)}{\mathsf{Pr}(s|\mathbf{M})}\log\left(\frac{\mathsf{Pr}(s|\mathbf{M}_i)}{\mathsf{Pr}(s|\mathbf{M})}\right),
\end{equation}
\normalsize

\noindent{}in which $\bigcup_{\mathbf{M}_i\in\mathcal{M}}=\mathbf{M}$. The partition $\mathcal{M}$ can be created using one or multiple member attributes. By combining multiple attributes, the model can detect subtle skill supply differences. For example, by grouping members using both industry and job title, we can discover that although skill \textsf{KDB+}, a financial database, is not popular among either \textsf{software developer}s or in the \textsf{financial industry}, it is a preferred skill in cohort \textsf{software developers} \& \textsf{financial industry}.

\noindent{\textbf{Job Features}:} Because demand implicitly influences the supply, here we also measure the skills' demand in terms of the job-side skill popularity $\mathsf{Pr}(s|\mathbf{P})$.
{Similar to member features, here we use pointwise mutual information $\mathsf{PMI}(l;s)$ to model the job-side skill popularity, where $l$ is the job posting segment label, \textit{e.g.}, summary and requirement.}

\subsection{\model}

After describing both salience and market-aware features, now we will discuss how we train the proposed \model ~model using the generated features. Recall that we need to learn the utility function defined in Def.~\ref{def:market_aware} to infer salience and market-aware skills for job targeting. With the job-skill pairs $\mathbf{J}^+$ and $\mathbf{J}^-$ that we collected from job posters and features described in Sec.~\ref{sec:job2skills}, we can learn the utility function $U$ by viewing this task as a binary classification problem where for a given skill $s$ and a job posting $p$, predict if $s$ is a salience and market-aware job targeting skill for $p$. 

Among all the possible machine learning models ranging from the generalized linear model to neural networks~\cite{cheng2016wide,lou2017bdt,he2017neural,zhang2016glmix}, here we chose to use XGBoost~\cite{chen2016xgboost} because it is fast, interpretable, and has good scalability with small memory footprint. By leveraging an in-house implementation of XGBoost, we were able to serve the model online without noticeable latency increase over the existing linear production model. The XGBoost-based \model ~is trained with a logistic regression loss to optimize the binary classification task, and we use the resulting tree-based \model ~model as the utility function $\mathsf{U}$ to extract market-aware job targeting skills for job postings. We define the loss function of the \model ~as:

\small
\begin{equation}
\begin{split}
    \mathcal{L} =& -\sum_{\langle p,s\rangle\in\mathbf{J}^+}\log\left(\sum_{k}f_k\left(\phi\left(s, c_p, \mathbf{M}, \mathbf{P}, \mathbf{S}_{c_p}\right)\right)\right) \\
    &- \sum_{\langle p,s\rangle\in\mathbf{J}^-}\log\left(1 - \sum_{k}f_k\left(\phi\left(s, c_p, \mathbf{M}, \mathbf{P}, \mathbf{S}_{c_p}\right)\right)\right) + \sum_{k}\Omega(f_k)
\end{split}
\end{equation}
\normalsize

\begin{table*}[t]
    \caption{Relative skill extraction AUROC improvement on JT and QA datsets.}
    \vspace{-.2cm}
    \label{tab:offline_skill_prediction}
    \centering
    \begin{adjustbox}{max width=\linewidth}
        \begin{tabular}{l|c c c}
        \toprule
            Model & Job Targeting Skills (JT) & Quality Applicant Skills (QA) & Overall (JT + QA) \\ 
         \midrule
           Salience\&Market-agnostic baseline & $-$ & $-$ & $-$ \\
           \model\ (trained w/ JT)  & $\mathbf{+55.77}\%$ & $+49.27\%$ & $+52.52\%$ \\
          \model\ (trained w/ QA)  & $+29.30\%$ & $\mathbf{+76.37}\%$ & $+52.84\%$ \\
         \textbf{\model\ (trained w/ JT+QA)}  & $+50.81\%$ & $+74.52\%$ & $\mathbf{+62.67}\%$ \\
       \bottomrule
        \end{tabular}
    \end{adjustbox}
     \vspace{-.2cm}
\end{table*}

\noindent{}in which $\mathbf{J}^+$ and $\mathbf{J}^-$ are the positive and negative job-skill pairs, $\phi(\cdot)$ denotes the combined market and salience feature vector of a given $\langle p,s \rangle$ pair, $f_k$ represents the $k^\mathsf{th}$ tree in the model, and $\Omega(f_k) = \gamma T + \frac{1}{2}\lambda||\mathbf{w}||^2$ is the regularization term that penalizes the complexity of tree $f_k$, in which $T$ denotes the number of leaves in tree $f_k$, $\mathbf{w}$ is the leaf weights, $\gamma$ and $\lambda$ are the regularization parameters.

\section{Experiments}~\label{sec:experiments}

In this section, we conduct an extensive set of experiments with both offline and online A/B tests to demonstrate the effectiveness of the proposed \model ~model compared to our market-agnostic production model. We also present a case study to demonstrate the actual skills returned by \model ~and how we can get better market insights from it.

The \model ~model evaluated in this section contain all aforementioned market-aware and multi-resolution skill salience features. Note that the job-level salience sub-model we used in production \model ~model is a FastText-based model instead of the BERT model we tried offline. This is because we observed significant latency reduction with only $3\%$ salience accuracy drop. 

The market-agnostic production model (baseline for short) we compared against is a logistic regression model trained with skill appearance features, e.g. job-level features such as \textit{is the skill mentioned in the text?}, \textit{where the skill is mentioned?}, and global-level features such as mention frequency. 

The offline training and evaluation data are collected using the following procedure. 
We used $16$ months of LinkedIn's English Premium jobs posted on LinkedIn as input and generated around $3$ million job-skill pairs for training and evaluation.
To be specific, we used the methods described in Sec.~\ref{subsec:data_collection} and collected 1) job-targeting (JT) dataset using job poster provided job targeting skills, and 2) quality applicant (QA) dataset using the common skills shared by job applicants who received positive feedbacks from recruiters.
We used $60\%$ of them for training, $20\%$ for validation, and the rest for testing. Note that unlike~\model, the production baseline is trained on the JT data only. During inference, both methods use the same skill tagger to get the same set of skill candidates from jobs.

\subsection{Offline Evaluation}

We present the~\model~offline evaluation result on the hold-out sets of the two training datasets (JT and QA), and report the relative AUROC improvement against the production baseline model. As shown in Tab.~\ref{tab:offline_skill_prediction}, \model ~significantly outperforms baseline by $55.77\%$ on the job targeting (JT) set and $49.27\%$ on the quality applicants' (QA) skill set. Moreover, by training with both human-labeled JT and derived QA dataset, we are able to generalize the model and achieve a better overall AUROC on both tasks by increasing the overall AUROC by $+62.67\%$.

\begin{table}[t]
    \caption{Feature ablation test on job targeting skill inference.}
    \vspace{-.2cm}
    \label{tab:feature_ablation}
    \centering
    \begin{adjustbox}{max width=\linewidth}
        \begin{tabular}{l|c}
            \toprule
            Model & AUROC Improvement ($\%$)\\
            \midrule
            Salience \& Market-agnostic baseline & $-$ \\
            \textbf{\model~w/ Market and Salience features} & $\mathbf{+56.99}\%$ \\ 
            \ \ \ -- Salience features only & $+55.43\%$ \\

            \ \ \ -- Market features (member+job) only & $+54.91\%$\\
            \ \ \ \ \ -- Member features only & $+46.47\%$\\
            \ \ \ \ \ -- Job features only & $+49.88\%$\\
            \bottomrule
        \end{tabular}
    \end{adjustbox}
    \vspace{-.2cm}
\end{table}
\begin{table}[t]
    \centering
    \caption{Online A/B test result on the LinkedIn Job Recommendation (JYMBII~\cite{kenthapadi2017personalized}) page.}
    \vspace{-.2cm}
    \label{tab:online_jymbii_test}
      \begin{adjustbox}{max width=\linewidth}

    \begin{tabular}{l|c c c}
    \toprule
           & Onsite Apply & Job Save & Member Coverage  \\
    \midrule
    \model & $+1.92\%$ & $+2.66\%$ & $+6.71\%$ \\
    \bottomrule
    \end{tabular}
    \end{adjustbox}
         \vspace{-.2cm}
\end{table}

Next, we present an ablation study to learn the importance of each feature group and present the relative AUROC improvement in Tab.~\ref{tab:feature_ablation}. Note the evaluation dataset used here is a slightly different JT dataset collected using the same procedure as Tab.~\ref{tab:offline_skill_prediction} but different time span. We can see both salience and market feature group positively contribute to the model performance improvement. We also observe that when only using one feature group, model trained with deep learning-based salience feature is $0.95\%$ better than market-feature only model. By combining both group of features, we further improve the AUROC by $1.56\%$ comparing to the salience only model. These improvements indicate that market-dynamics modeled by market-aware features provide additional information on the skill importance for job targeting that cannot be captured by modeling job posting-based skill salience only.

In addition to the ablation study, we also looked at the feature importance of the \model~model trained using both market and salience features. We found that all three salience features are ranked within the top-$5\%$ most important features, and segment-level salience feature is the most important one followed by the sentence-level and job-level salience features. This means both deep learning powered salience features and market features are crucial to the model and cannot be replaced by each other.

\subsection{Online Job Recommendations}

In this section, we deploy \model~to production, apply it to all LinkedIn Jobs to extract skill entities for job targeting, and retrain our job recommender system, \textit{Jobs You May Be Interested In} (JYMBII~\cite{kenthapadi2017personalized}), based on the extracted salience and market-aware job targeting skills. We perform online A/B test with $20\%$ of the LinkedIn traffic for $7$ days, and observe significant lift in multiple metrics. As shown in Tab.~\ref{tab:online_jymbii_test}, the \model-based JYMBII model not only recommends better jobs (reflected by increased job apply and save rate), but also increases the percentage of members receive job recommendations.

\begin{table}[t]
    \centering
    \caption{A/B test result of job targeting skill suggestions.}
    \vspace{-.2cm}
    \label{tab:online_test}
      \begin{adjustbox}{max width=\linewidth}

    \begin{tabular}{l|c c c c}
    \toprule
    Model & Skill Add Rate & Skill Reject Rate \\
    \midrule
    \model\ (market-aware)& $-31.44\%$ & $-33.71\%$ \\
    \model\ (market- and salience-aware) & $\mathbf{-33.75}\%$ & $\mathbf{-37.06}\%$\\
    \bottomrule
    \end{tabular}
    \end{adjustbox}
    \vspace{-.4cm}
\end{table}

\begin{table*}[!t]
  \caption{Sample of top-$10$ job targeting skills extracted by the salience and market-agnostic baseline and \model.}
\vspace{-.2cm}
  \label{tab:top_skill_by_title}
  \begin{adjustbox}{max width=.9\linewidth}
\begin{tabular}{l|l|l|l}
\toprule
Job Title & Shared by Both Models & 
Skills returned by \model~only & Skills returned by baseline only  \\
\midrule
Software Engineer & - & \begin{tabular}[c]{@{}l@{}}Design, Java, Communication, C, C++, Management, \\ Javascript, SQL, Cloud Computing, Architecture\end{tabular} & \begin{tabular}[c]{@{}l@{}}OpenGL, DropWizard, ActiveRecord, LLVM, \\ Sinatra, C++0x, Guice, GRPC, cmake, Boost C++\end{tabular} \\
\midrule
Data Scientist & \begin{tabular}[c]{@{}l@{}}Data Science,\\ Machine Learning\end{tabular} & \begin{tabular}[c]{@{}l@{}}Data Mining, Python, Analytics, Pattern Recognition, \\ Statistics, R, AI, Communication\end{tabular} & \begin{tabular}[c]{@{}l@{}}Apache Spark, Predictive Modeling, Statistical Modeling, \\ scikit-learn, Deep Learning, Text Mining, Pandas, Keras\end{tabular} \\
\midrule
Audit Tax Manager & Auditing & \begin{tabular}[c]{@{}l@{}}Communication, Management, Research, Tax Preparation, \\ Supervisory Skills, Presentations, Tax Compliance, Engagements, Budgeting\end{tabular} & \begin{tabular}[c]{@{}l@{}}Tax, Due Diligence, US GAAP, Financial Accounting,  \\ Financial Audits, External Audit, GAAP, IFRS, Internal Controls\end{tabular}\\
\bottomrule
\end{tabular}
  \end{adjustbox}
      \vspace{-.2cm}
\end{table*}

\begin{table}[t]
  \caption{Sample of top-$5$ job targeting skills per industry and country extracted by the salience and market-agnostic baseline and \model. Red squared skills are sub-optimal.}
      \vspace{-.2cm}
  \label{tab:us_in_skills}
    \begin{adjustbox}{max width=\linewidth}

\begin{tabular}{c|cc|cc}
\toprule
\multicolumn{1}{l}{} \vline& \multicolumn{2}{c}{United States} \vline & \multicolumn{2}{c}{India} \\
\multicolumn{1}{c}{Industry} \vline & \multicolumn{1}{c}{Baseline} & \multicolumn{1}{c}{\model} \vline & \multicolumn{1}{c}{Baseline} & \multicolumn{1}{c}{\model} \\
\midrule
\multirow{5}{*}{Government} & \colorbox{red!20}{teaching} & analytical skills & start-ups & communication \\
 & \colorbox{red!20}{TSO} & communication & management & think tank \\
 & management & {army} & procurement & research \\
 & fire control & {defense} & finance & social media \\
 & law enforcement & hazardous materials & human resources & disability rights \\\midrule
\multirow{5}{*}{Technology} & management & analytical skills & SAP products & customer experience \\
 & sales & project management & management & analytical skills \\
 & cloud computing & communication & \colorbox{red!20}{Jakarta EE} & solution architecture \\
 & consulting & sales & cloud computing & business process \\
 & \colorbox{red!20}{salesforce.com} & problem solving & \colorbox{red!20}{salesforce.com} & helping clients\\
 \bottomrule
\end{tabular}
\end{adjustbox}
  \vspace{-.4cm}
\end{table}

\subsection{Online Job Targeting Skill Suggestions}

We also apply \model ~to provide job targeting skill suggestions in LinkedIn's job posting flow. When a recruiter posts a job on LinkedIn, we use \model ~to recommend $10$ skills, and the poster will be able to save at most $10$ job targeting skills by either selecting from the recommendation or providing their own. We ramped our \model\  model to $50\%$ of LinkedIn's traffic, and report the $4$-week A/B test result. The definition of the metrics are as follows:

\begin{itemize}[leftmargin=*]
\item \textbf{skill add rate}: \% of skills that are not recommended and are manually added,
\item \textbf{skill reject rate}: \% of recommended skills that are rejected. 
\end{itemize}

As shown in Tab.~\ref{tab:online_test}, the skills recommended by \model ~are notably better than the existing production model because the recruiters are now $31.44\%$ less likely to manually add a job targeting skill and $33.71\%$ less likely to reject recommendation. In general, the \model ~increases the overall job targeting skill coverage and quality by adjusting skill importance as a function of the skills' salience and market signals.

\section{Qualitative Analysis}
We have shown \model~ improves job targeting at LinkedIn because it captures salience and market supplies. Here we further demonstrate salience and market-aware skills' strength in market analysis. We claim that \model~ can capture hiring trend --- \emph{e.g.,} \textit{what skill sets are required in different sectors?} \textit{what kinds of talents employers are recruiting?} --- very well. As we explained, \model~ is trained on the hiring experts feedback and market supplies. In other words, it is trained on signals representing what kinds of people (skill sets) the companies want the most, and what kinds of people that actually got hired. Next we show that \model's results can reveal the intent of employers better than the baseline model which based on named entity recognition.

\subsection{Skill Trend Insight}

\begin{figure}[t] 
\centering
  \begin{subfigure}[b]{\linewidth}
    \centering
    \includegraphics[width=\textwidth]{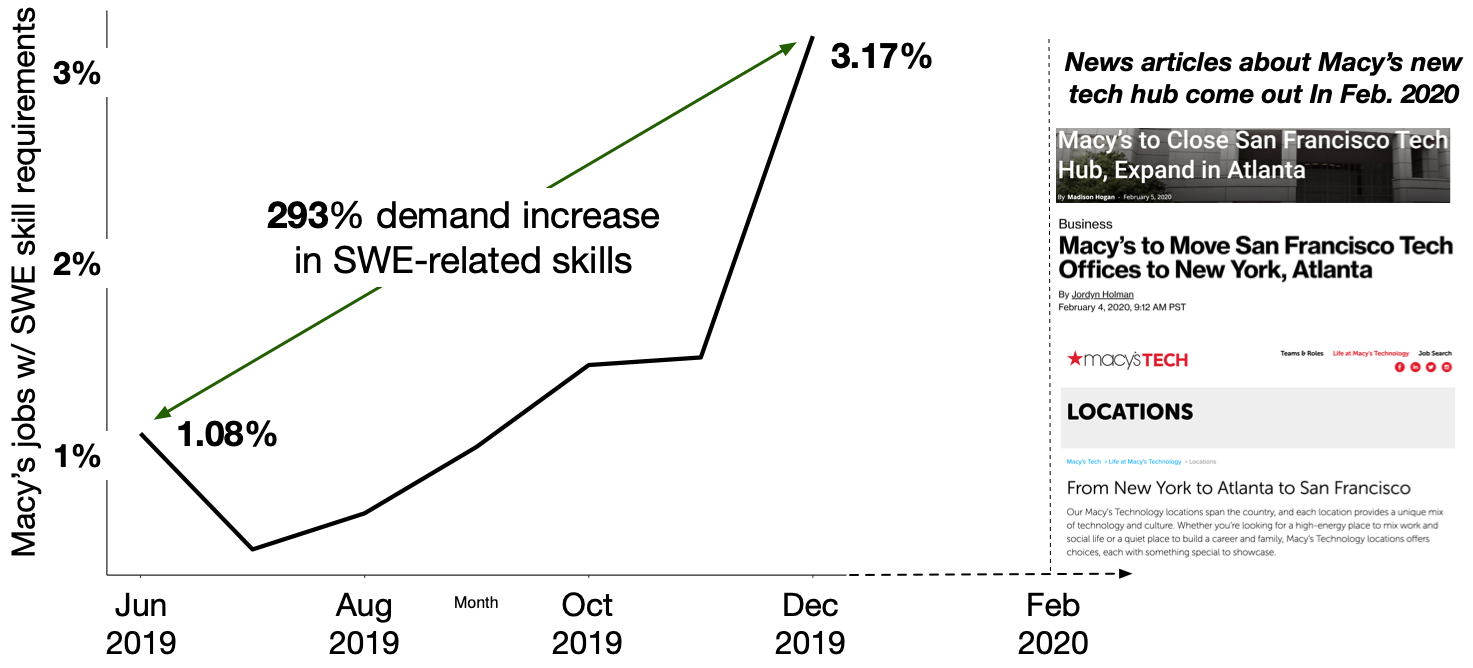}
  \end{subfigure}%
  \caption{\% of Macy's jobs posted per month from Jun. to Dec. 2019 that require software development related skills.} 
    \vspace{-.3cm}
\label{fig:macy_trend} 
\end{figure}

\begin{figure}[t] 
\centering
  \begin{subfigure}[b]{\linewidth}
    \centering
    \includegraphics[width=\textwidth]{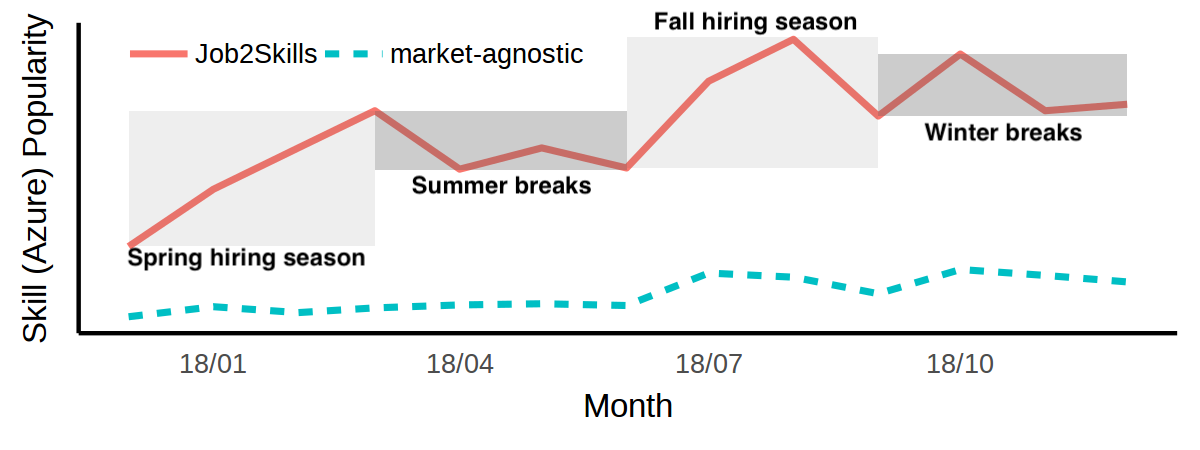}
  \end{subfigure}%
  \vspace{-.3cm}
  \caption{Azure's skill popularity in 2018.} 
    \vspace{-.2cm}
\label{fig:skill_insight} 
\end{figure}

The first kind of cases is about the trend of skills. \textit{Which skills are getting popular? Which company is growing a certain job segment?} We show that \model's results can be used to forecast this trend even before actual news articles coming out.

In 2019, we identified Macy's tech expansion using the results of \model. In Fig.~\ref{fig:macy_trend}, we presented the percentage of Macy's monthly posted jobs that require skills such as \textsf{Java}, \textsf{Javascript}, and \textsf{SQL}. We found Macy's demand on tech skills almost tripled from June to December, 2019. We suspect such radical change indicates Macy's is planning to invest into technology section. Two months after we detected this trend, in February, 2020, Macy's officially announced its tech operation expansion in Atlanta and New York\footnote{https://www.bloomberg.com/news/articles/2020-02-04/macy-s-to-move-san-francisco-tech-offices-to-new-york-atlanta}.

Besides predicting company expansions, \model ~also better captured the skill supply and demand regarding market dynamics such as recruiting circles. We present the skill popularity trend of \textsf{Azure}, the fifth trending skill in SF Bay Area, in Fig.~\ref{fig:skill_insight}. We found that job targeting skill popularity is highly correlated with major hiring and vacation seasons. Interestingly, \model ~popularity changes also correlated with other market signals such as company performance. The first gray box in Fig.~\ref{fig:skill_insight} highlights a significant four-month increase trending from Dec. 2017 to Mar. 2018 of \textsf{Azure}, which aligns with MSFT's 2018Q3 earning that shows Azure cloud has a $93\%$ revenue growth. We suspect the correlation is caused by the adoption of Azure services in the market -- more companies are using Azure hence the skill becomes more popular, and such market share increase also reflect in the revenue growth.

\subsection{Skill Insights}

The second kind of cases is revealing diversified skill demands in different industry sections and regions. We show that the job targeting skills generated by \model ~better captures such market diversity by modeling salience and market signals jointly.

In Tab.~\ref{tab:top_skill_by_title} we present the top-$10$ job targeting skills of three occupations, \textsf{Software Engineer}, \textsf{Data Scientist}, and \textsf{Audit Tax Manager}. It is clear that unlike the salience and market-agnostic model which mainly focuses on very specific skills with limited supplies such as C++0x and DropWizard, \model ~is able to return a diversified set of skills at the right granularity ranging from popular programming languages to soft skills such as communication and management.

Besides better representing skills for different occupations, the skills generated by \model can also capture skill supply and demand differences in different industries and regions. Here we compared top-$5$ job targeting skills between government and technology industries in the United States and India. As shown in Tab.~\ref{tab:us_in_skills}, skills returned by the proposed \model~ is significantly different from the market and salience agnostic baseline model. we believe this is because \model~ considers skill salience and market dynamics. For example, the baseline model wrongly pick \textsf{TSO} (Time Sharing Option) as a top skill in US government industry because it is mentioned in many jobs posted by \textsf{Transportation Security Administration} and \textsf{The Department of Homeland Security}. However, the TSO mentioned in those jobs actually refers to \textsf{Transportation Security Officer}. By evaluating the skill salience, \model~ was able to identify TSO is not a valid skill. In addition, the market and salience agnostic baseline also ranked many specific tools such as \textsf{SAP} and \textsf{Jakarta EE} as top skills in Indian technology industry. Instead of targeting on very specific skills, the proposed \model~ measures the supply in the Indian market and selected many job-related, high supply skills such as \textsf{customer experience} and \textsf{helping clients}.

The skills extracted by \model~ can reveal market insights due to its market-awareness. It is interesting that the government industry market of US is significantly different from India. While India has a focus on a variety of skills ranging from research to social media to rights, US mostly focuses on military related skills. We suspect this is because US government positions are mostly defense/environmental related positions and prefer veterans. The technology industry is quite different between US and India, too. As shown in Tab.~\ref{tab:us_in_skills}, it is clear that the Indian market has a focus on IT support whereas US is more about management and sales.

\begin{table}[t]
\caption{Top-$5$ job targeting skills of US government and technology generated the salience and market-agnostic baseline and \model. Red squared skills are sub-optimal.}
      \vspace{-.2cm}

\label{tab:skill_senority}
  \begin{adjustbox}{max width=\linewidth}
\begin{tabular}{r|cc | cc}
\toprule
 & \multicolumn{2}{c}{Director} & \multicolumn{2}{c}{Entry} \\
Industry & Baseline & \model & Baseline & \model \\\midrule
\multirow{5}{*}{Government} & \colorbox{red!20}{DES} & leadership & \colorbox{red!20}{TSO} & communication \\
 & management & analytical skills & \colorbox{red!20}{DES} & defense \\
 & \colorbox{red!20}{teaching} & project management & fire control & hazardous materials \\
 & communication & management & \colorbox{red!20}{teaching} & analytical skills \\
 & leadership & interpersonal skills & management & army \\
 \midrule
\multirow{5}{*}{Technology} & sales & analytical skills & \colorbox{red!20}{sales} & analytical skills \\
 & management & consulting & cloud computing & communication \\
 & cloud computing & project management & devops & SQL \\
 & leadership & sales & \colorbox{red!20}{DES} & software development \\
 & \colorbox{red!20}{DES} & communication & \colorbox{red!20}{management} & Java \\\bottomrule
\end{tabular}
\end{adjustbox}
  \vspace{-.4cm}
\end{table}

Moreover, we found that \model ~captures the skill differences between seniority levels. In Tab.~\ref{tab:skill_senority}, we presented the top-$5$ job targeting skills of US government and technology industries generated by the baseline and the proposed \model. Compared to the baseline, skills generated by \model~ are more representative and capture the skill shift across different seniority levels. For example, entry level positions require domain-specific skills such as \textsf{SQL} and \textsf{Java}, and higher level roles, regardless the industry, focus more in management skills such as \textsf{management} and \textsf{leadership}. The baseline model also selects many less relevant skills such as \textsf{DES (Data Encryption Standard)} and \textsf{management} for entry-level technology role due to the lack of salience and market modeling.

\section{Related Work}~\label{sec:related_work}

\noindent{\textbf{Job Recommendation}}.
Previous work usually treat the job targeting problem as job recommendation~\cite{volkovs2017content, guo2017integration}, and optimizes the model using direct user interaction signals such as click, dismiss, bookmark, etc.~\cite{agarwal2009spatio,abel2017recsys,huang2019online,matn2020behavioral}. Borisyuk~\textit{et al.} proposed LiJar~\cite{borisyuk2017lijar} to redistribute job targeting audiences and improve marketplace efficiency. Dave~\textit{et al.} designed a representation-learning method to perform job and skill recommendations~\cite{dave2018combined} \hl{. Li~\textit{et al.} used career history to predict next position~\cite{li2017nemo}.} None of the previous works address the most pressing job targeting issue, which is how to properly represent jobs with relevant, important attribute entities to improve the number of quality applicants a job can reach.

\noindent{\textbf{Skill Analysis}}.
Traditionally, skill analysis are often conducted by experts manually to either gain insights~\cite{prabhakar2005skills} or curate structured taxonomy~\cite{national2010database}. Recently, SPTM~\cite{xu2018measuring} used topic modeling to measure the popularity of $1,729$ IT skills from $892,454$ jobs. TATF~\cite{wu2019trend} is a trend-aware tensor factorization method that models time-aware skill popularity. DuerQuiz~\cite{qin_duerquiz:_2019} is proposed to create in-depth skill assessment questions for applicant evaluation. These methods were applied to small-sacle IT jobs only and are not designed to extract skills for job targeting purpose. Recently Xiao proposed a social signal-based method for members' skill validation~\cite{yan2019social}. However it is not applicable to jobs due to the lack of such signals.

\noindent{\textbf{Job Market Analysis}}.
Modeling job targeting and recommendation using skills is mostly inspired by economic research which analyzes the labor market using skills as the most direct and vital signal~\cite{autor2003skill,saar2017participation}. However these works are either conducted on a very small scale or using only a handful of hand-crafted general skill categories. Woon~\textit{et al.}~\cite{woon2015changes} performed a case study to learn occupational skill changes, but the skills are limited to $35$ skills provided by O*NET~\cite{peterson2001understanding}. Radermacher~\textit{et al.}~\cite{radermacher2014investigating} studied the skill gap between fresh graduates and industry expectations based on the feedback of $23$ managers and hiring personnel using $16$ hand picked skills. Recently, ~\cite{vasudevan2018estimating,johnston2017estimating} analyzed labor demand and skill fungibility using a skill taxonomy with $1,351$ skills in the IT industry. APJFNN~\cite{qin2018enhancing} and other resume-based method~\cite{zhu2018person} are developed to predict person-job fit by comparing the job description and resume. HIPO~\cite{ye_identifying_2019} identifies high potential talent by conducting neural network-based social profiling. OSCN~\cite{sun_impact_2019} and HCPNN~\cite{meng_hierarchical_2019}, use recurrent neural networks and attention mechanism to predict organization and individual level job mobility. However, none of these works addresses the market-aware job targeting task, and they all use a limited skill taxonomy that contains at most a thousand skills.

\section{Conclusion and Future Work}\label{sec:conclusion}

In this work, we proposed salience and market-aware skill extraction task, discussed two data collection strategies, and presented \model, which models skill salience using deep learning methods and market supply signals using engineered features. Lastly, we conducted extensive experiments and showed that \model ~significantly improves the quality of multiple LinkedIn products including job targeting skill suggestions and job recommendation. We also performed large-scale case studies to explore interesting insights we obtained by analyzing \model ~results. In future work, we plan to add temporal information into the model \hl{and explore advanced methods to learn skill embeddings}.

\bibliographystyle{ACM-Reference-Format}
\bibliography{bibliography}

\end{document}